\documentclass[12pt]{article}

\usepackage{a4}

\usepackage{bm}

\usepackage{amsmath}
\usepackage{amsfonts}

\newcommand{\tr}{{\rm tr}}
\newcommand{\ol}{\overline}

\newcommand{\wh}{\widehat}

\newcommand{\rap}[2]
{\setbox1=\hbox{#1}%
\setbox2=\hbox to\wd1{\hss #2\hss}%
\mbox{\rlap{\box1}\box2}}

\newcommand{\sla}[1]{\rap{$#1$}{$\backslash$}}

\newcommand{\GG}{{\Gamma_{\rm iso}}}

\begin{document}
\begin{titlepage}
\title{
\begin{flushright}
\normalsize{TIT/HEP-620\\
Sep 2012}
\end{flushright}
       \vspace{2cm}
Supersymmetric theories on squashed five-sphere
       \vspace{2cm}}
\author{
Yosuke Imamura\thanks{E-mail: \tt imamura@phys.titech.ac.jp}$^{~1}$
\\[30pt]
{\it $^1$ Department of Physics, Tokyo Institute of Technology,}\\
{\it Tokyo 152-8551, Japan}
}
\date{}

\maketitle
\thispagestyle{empty}

\vspace{0cm}

\begin{abstract}
We construct supersymmetric theories on the $SU(3)\times U(1)$ symmetric
squashed five-sphere with $2$, $4$, $6$, and $12$ supercharges.
We first determine the Killing equation by dimensional
reduction from 6d,
and use Noether procedure to construct actions.
The supersymmetric Yang-Mills action is straightforwardly
obtained from the supersymmetric Chern-Simons action
by using a supersymmetry preserving constant vector multiplet.
\end{abstract}

\end{titlepage}

\section{Introduction}
Recently five-dimensional (5d) supersymmetric (SUSY) gauge theories have
attracted much interest.
It is known that there exist 5d gauge theories with non-trivial fixed points\cite{Seiberg:1996bd}, and their dynamics is closely related to the brane physics
in string theory.
The relation to 
six-dimensional ${\cal N}=(2,0)$ theories provides
another motivation to study 5d theories.
There is an interesting proposal\cite{Douglas:2010iu,Lambert:2010iw}
that a $(2,0)$ theory on a 6d manifold
${\cal M}\times{\bm S}^1$
be equivalent to a 5d SUSY
Yang-Mills theory on ${\cal M}$.
Because we have no Lagrangian description of $(2,0)$ theories
and cannot directly analyze them,
this relation provides an important access to $(2,0)$ theories through 5d
gauge theories.

The first step to analyze a theory is
to construct the action.
The action of the ${\cal N}=1$ SUSY theory
on the flat ${\bm R}^5$
is given in \cite{Seiberg:1996bd}.
SUSY gauge theories on various curved backgrounds are
also used to obtain exact results by localization.
Theories on
the round ${\bm S}^5$\cite{Hosomichi:2012ek,Kim:2012av},
${\bm S}^4\times{\bm S}^1$\cite{Kim:2012gu,Terashima:2012ra},
and ${\bm S}^3\times{\bm R}^2$\cite{Kawano:2012up}
have been constructed.
SUSY theories
on contact manifolds are constructed
in \cite{Kallen:2012cs}.
The perturbative part of the ${\bm S}^5$ partition function
for the round ${\bm S}^5$
is computed
in \cite{Kallen:2012cs,Kallen:2012va,Kim:2012av},
and used to confirm
predictions of
AdS/CFT correspondence\cite{Kim:2012av,Kallen:2012zn,Jafferis:2012iv}.
The superconformal index is computed
in \cite{Kim:2012gu} for ${\cal N}=1$ SUSY gauge theories and
the symmetry enhancement at the strong coupling limit is
investigated.

The purpose of this paper is to give more examples
of 5d SUSY theories on a curved background.
We construct SUSY actions on the $SU(3)\times U(1)$
symmetric squashed ${\bm S}^5$ with
the metric
\begin{equation}
ds_{{\bm S}^5}^2=ds_{\bm{CP}^2}^2+\frac{1}{v^2}(d\psi+V)^2.
\label{squashedmetric}
\end{equation}
We treat ${\bm S}^5$ as a Hopf fibration over $\bm{CP}^2$.
The first and the second terms in 
(\ref{squashedmetric}) are the metric of the base $\bm{CP}^2$ and
that of the Hopf fiber, respectively.
They are normalized so that when $v=1$
(\ref{squashedmetric}) gives
the round ${\bm S}^5$ with radius $r$.

The eight supercharges of ${\cal N}=1$ SUSY on the round ${\bm S}^5$
belong to ${\bm 4}+\ol{\bm 4}$ of the isometry group $SO(6)_{\rm iso}$.
The squashing breaks $SO(6)_{\rm iso}$ to $SU(3)\times U(1)$.
Correspondingly, the supercharges split to
${\bm 3}_{+1}+\ol{\bm3}_{-1}$ and ${\bm1}_{-3}+{\bm1}_{+3}$.
We will show that only one of ${\bm 3}_{+1}+\ol{\bm3}_{-1}$
or ${\bm1}_{-3}+{\bm1}_{+3}$ can be preserved in the squashing.
We call these two kinds of preserved SUSY
${\cal N}=3/4$ and ${\cal N}=1/4$.
We also construct theories with the number of supercharges doubled,
${\cal N}=3/2$ and ${\cal N}=1/2$ theories,
by combining a vector multiplet and an adjoint hypermultiplet
with a critical value of the mass parameter.

The organization of this paper is as follows.
In the next section, we summarize SUSY gauge theories
on conformally flat backgrounds.
We explain how we can obtain a SUSY Yang-Mills theory on the
round ${\bm S}^5$ from a SUSY Chern-Simons theory
in the same background,
which can be obtained from the theory on the flat ${\bm R}^5$
by using Weyl transformation.
We also review the relation between SUSY in 5d and that in 6d
following \cite{Kim:2012av}.
In \S\ref{squashing.sec} we construct SUSY theories on the
squashed five-sphere.
We first determine the Killing equation by using
a twisted compactification of ${\bm S}^5\times{\bm R}$,
and construct SUSY actions
by Noether procedure.
We also construct ${\cal N}=1/2$ and ${\cal N}=3/2$ theories by
combining a vector multiplet and an adjoint hypermultiplet.
\S4 is devoted to discussion.
Conventions for $SU(2)$ and $SO(5)$ are summarized in the appendix.

We use $\mu,\nu,\ldots$ for $5d$ vector indices.
We use local orthonormal frame unless otherwise noted.
$I,J,\ldots$ and $a,b,\ldots$ are used for $SU(2)_R$-doublet and triplet
indices, respectively, and $A,B,\ldots$ for
$SU(2)_F$-doublet indices.
For more details see the appendix.
\section{${\cal N}=1$ in conformally flat backgrounds}
\subsection{Conformal theories in 5d}
Let us first consider vector multiplets ${\cal V}^\alpha$ of ${\cal N}=1$
SUSY gauge theories in the flat ${\bm R}^5$.
A vector multiplet ${\cal V}^\alpha$
consists of
a gauge field $A_\mu^\alpha$,
a real scalar field $\phi^\alpha$,
a symplectic Majorana spinor $\lambda_I^\alpha$ ($I=1,2$),
and three real auxiliary fields $D_a^\alpha$ ($a=1,2,3$).
$\lambda_I^\alpha$ and $D_a^\alpha$ form an $SU(2)_R$ doublet and a triplet, respectively.
The action of vector multiplets is specified by
the prepotential ${\cal F}(\phi)$,
a gauge invariant real function of $\phi^\alpha$.
The Lagrangian density on the flat ${\bm R}^5$ is\cite{Seiberg:1996bd}
\begin{align}
{\cal L}_0^{\rm vector}
&=
{\cal F}_{\alpha\beta}\bigg(
\frac{1}{4}F^\alpha_{\mu\nu}F^{\beta\mu\nu}
+\frac{1}{2}D_\mu\phi^{\alpha} D^\mu\phi^{\beta}
-\frac{1}{2}D_a^\alpha D_a^\beta
+\frac{1}{2}\epsilon^{IJ}(\lambda^{\alpha}_I\sla D\lambda^{\beta}_J)
-\frac{1}{2}\epsilon^{IJ}(\lambda_I^\alpha[\phi,\lambda_J]^\beta)
\bigg)
\nonumber\\&
+{\cal F}_{\alpha\beta\gamma}\left[
\left(\frac{i}{24}\epsilon^{\lambda\mu\nu\rho\sigma}A_\lambda^\alpha F^\beta_{\mu\nu}F^\gamma_{\rho\sigma}+\cdots\right)
-\frac{i}{4}\epsilon^{IJ}(\lambda_I^\alpha\sla F^\beta\lambda_J^\gamma)
+\frac{1}{4}\epsilon^{IK}(\tau_a)_K{}^J D_a^\alpha(\lambda_I^\beta\lambda_J^\gamma)\right],
\label{fiveprep}
\end{align}
where
${\cal F}_{\alpha\beta}$ and ${\cal F}_{\alpha\beta\gamma}$ are defined by
\begin{equation}
{\cal F}_{\alpha\beta}
=
\frac{\partial}{\partial\phi^\alpha}
\frac{\partial}{\partial\phi^\beta}
{\cal F}(\phi),\quad
{\cal F}_{\alpha\beta\gamma}
=
\frac{\partial}{\partial\phi^\alpha}
\frac{\partial}{\partial\phi^\beta}
\frac{\partial}{\partial\phi^\gamma}
{\cal F}(\phi).
\end{equation}
The gauge covariant derivative and the field strength
are defined by
\begin{equation}
D=d-i[A,*],\quad
F=dA-iA\wedge A.
\end{equation}
When we consider a curved background
$D_\mu$ also contain the spin connection.
The terms in the parenthesis in the second line
of (\ref{fiveprep})
is the Chern-Simons action.
The dots represent terms with one or no derivative,
which exist when the gauge group is non-Abelian.
The gauge invariance
requires the coefficients ${\cal F}_{\alpha\beta\gamma}$
of the Chern-Simons term to be constant.
This means that the prepotential is an at most cubic polynomial.
The constant and linear terms in the prepotential do not affect the
action in the flat spacetime, while
we will see that the linear terms
give the Fayet-Iliopoulos action in curved backgrounds.

A superconformal theory
on a conformally flat background can be easily obtained by
Weyl transformation from the theory on the flat ${\bm R}^5$.
This is pointed out for 5d theories in \cite{Kim:2012gu}.
If the prepotential is
a cubic homogeneous polynomial,
the action (\ref{fiveprep})
is invariant under not only rigid SUSY
transformation
but also superconformal transformation
\begin{align}
\delta A_\mu&=\epsilon^{IJ}(\epsilon_I\gamma_\mu\lambda_J),
\nonumber\\
\delta \phi&=-i\epsilon^{IJ}(\epsilon_I\lambda_J),
\nonumber\\
\delta\lambda_I
&=-\frac{1}{2}\gamma^{\mu\nu}\epsilon_IF_{\mu\nu}
+i\gamma^\mu\epsilon_I D_\mu\phi
+iD_a(\tau_a)_I{}^J\epsilon_J
+2i\kappa_I\phi
,\nonumber\\
\delta D_a
&=-i\epsilon^{IK}(\tau_a)_K{}^J(\epsilon_I\gamma^\mu D_\mu\lambda_J)
+i\epsilon^{IK}(\tau_a)_K{}^J(\epsilon_I[\phi,\lambda_J])
+i\epsilon^{IK}(\tau_a)_K{}^J(\kappa_I\lambda_J),
\label{deltavector}
\end{align}
where the parameters $\epsilon_I$ and $\kappa_I$ are
symplectic Majorana
spinors satisfying
the Killing equation
\begin{equation}
D_\mu\epsilon_I=\gamma_\mu\kappa_I.
\label{killing5d}
\end{equation}
Furthermore,
We can make this action invariant under the local
Weyl transformation
\begin{equation}
g_{\mu\nu}=e^{-2\alpha}g'_{\mu\nu},\quad
A=A',\quad
\phi=e^\alpha\phi',\quad
\lambda=e^{\frac{3}{2}\alpha}\lambda',\quad
D=e^{2\alpha}D',
\label{vectorweyl}
\end{equation}
by introducing the curvature coupling of the scalar fields.
\begin{equation}
{\cal L}^{\rm vector}=
{\cal L}_0^{\rm vector}+
\frac{R}{4}{\cal F}(\phi).
\end{equation}
With the Weyl transformation
(\ref{vectorweyl}),
we can easily construct
the ${\cal N}=1$ SUSY
Chern-Simons action on conformally flat backgrounds.
The SUSY
Yang-Mills action and the Fayet-Iliopoulos action
are also easily constructed
with the help of a constant vector multiplet
as we will explain in the next subsection.

We use on-shell formalism for hypermultiplets.
A hypermultiplet consists of four real scalar fields $q_i$ ($i=1,2,3,4$)
and a symplectic Majorana spinor field $\psi_A$ ($A=1,2$).
The largest symmetry of
$k$ hypermultiplets is $SU(2)_R\times Sp(k)$,
and an arbitrary subgroup of $Sp(k)$ can be gauged.
We mainly focus only on the subgroup $SU(2)_F\times U(k)\subset Sp(k)$.
We write down actions and transformation laws
as if the gauge group $G$ is a subgroup of $U(k)$ and
hypermultiplets belong to the adjoint representation of $G$.
Extension to more general case is straightforward.
The kinetic action of hypermultiplets on the flat ${\bm R}^5$ is
\begin{align}
{\cal L}_0^{\rm hyper}
=&
\frac{1}{2}D_\mu q_i D^\mu q_i
-\frac{1}{2}\epsilon^{AB}(\psi_A\sla D\psi_B)
+\frac{1}{2}(\tau_a)_{ij}q_i[D_a,q_j]
+\frac{1}{2}[q_i,\phi][\phi ,q_i]
\nonumber\\&
+\epsilon^{AB}(\ol\rho_i)_A{}^I\psi_B[\lambda_I,q_i]
-\frac{1}{2}\epsilon^{AB}(\psi_A[\phi,\psi_B]),
\label{hyperaction}
\end{align}
where $(\rho_i)_I{}^A$ and $(\ol\rho_i)_A{}^I$ are
$SU(2)_R\times SU(2)_F$ invariant tensors, and
$(\tau_a)_{ij}$ is the 't Hooft symbol defined by
$(\tau_a)_{ij}=-(1/2)(\tau_a)_I{}^J(\rho_i)_J{}^A(\ol\rho_j)_A{}^I$.
This is invariant under the superconformal transformation
\begin{align}
\delta q_i&=-i\epsilon^{IJ}(\rho_i)_J{}^A(\epsilon_I\psi_A),\nonumber\\
\delta\psi_A&=
i(\ol\rho_i)_A{}^I\gamma^\mu\epsilon_ID_\mu q_i
+3i(\ol\rho_i)_A{}^I\kappa_Iq_i.
\end{align}
The Lagrangian
\begin{equation}
{\cal L}^{\rm hyper}={\cal L}^{\rm hyper}_0+\frac{3R}{32}q_iq_i
\end{equation}
improved by
the curvature coupling of the scalar fields
is invariant under the local Weyl transformation
\begin{equation}
g_{\mu\nu}=e^{-2\alpha}g'_{\mu\nu},\quad
q_i=e^{\frac{3}{2}\alpha}q_i',\quad
\psi=e^{2\alpha}\psi'.
\label{weylhyper}
\end{equation}
Note that the Weyl weights of fields in hypermultiplets
are protected by the superconformal algebra.
$q_i$ and $\psi_A$ have canonical weights $3/2$ and $2$, respectively.
We can use (\ref{weylhyper}) to obtain the action and the
transformation laws for hypermultiplets
in an arbitrary conformally flat background.

\subsection{Round ${\bm S}^5$}
The quadratic term in the prepotential
\begin{equation}
{\cal F}_{\rm YM}=\frac{1}{2g^2_{\rm YM}}\tr\phi^2
\label{ymprep}
\end{equation}
gives the Yang-Mills kinetic term
\begin{equation}
{\cal L}=\frac{1}{4g_{\rm YM}^2}\tr(F_{\mu\nu}F^{\mu\nu}).
\label{yangmillsterm}
\end{equation}
Although this is not conformal in 5d,
we can easily construct the SUSY Yang-Mills action
on a conformally flat background.
For concreteness and as a preparation for the next section,
let us consider the case of the round ${\bm S}^5$ with radius $r$.
On the round ${\bm S}^5$ the parameters $\epsilon$ and $\kappa$
(From this subsection we omit $SU(2)$ indices. See Appendix for the rules.)
belong to ${\bm 4}+\ol{\bm 4}$ of the isometry group $SO(6)_{\rm iso}$.
The spinors in each irreducible representation
satisfy
\begin{equation}
\kappa^{\bm4}=-\frac{i}{2r}\epsilon^{\bm4},\quad
\kappa^{\ol{\bm4}}=\frac{i}{2r}\epsilon^{\ol{\bm4}}.
\label{twoeq}
\end{equation}
It is convenient to define the chirality operator $\GG$ for $SO(6)_{\rm iso}$
which acts on ${\bm 4}$ and $\ol{\bm4}$ as $+1$ and $-1$, respectively.
We combine two equations in (\ref{twoeq}) into
\begin{equation}
\kappa=-\frac{i}{2r}\GG\epsilon.
\label{kappas5ande}
\end{equation}
With the relation (\ref{kappas5ande})
and the transformation laws in
(\ref{deltavector}),
we can show that the constant vector multiplet
\begin{equation}
{\cal V}^{(1)}=(\phi^{(1)},A^{(1)},\lambda^{(1)}_I,D_a^{(1)})
=
\left(1,0,0,\frac{i}{r}\delta_{a3}\right)
\label{constantround}
\end{equation}
preserves half of the supersymmetry
whose parameter satisfies
\begin{equation}
\tau_3\epsilon=\GG\epsilon.
\label{s5dircond}
\end{equation}
We can lift the prepotential ${\cal F}_{\rm YM}$
to a cubic polynomial by
multiplying $\phi^{(1)}=1$ to it.
Namely, we can obtain the SUSY Yang-Mills action
as a special SUSY Chern-Simons action
with the prepotential
\begin{equation}
{\cal F}=\phi^{(1)}{\cal F}_{\rm YM}=\frac{1}{2g_{\rm YM}^2}\phi^{(1)}\tr\phi^2.
\label{cubigym}
\end{equation}
In 5d the constant $1/g_{\rm YM}^2$ has mass dimension $1$,
and we can regard
(\ref{cubigym})
as a mass deformation to the Chern-Simons theory.
The supersymmetry preserved after such a mass deformation is
often called
rigid supersymmetry.
As in the case of ${\bm S}^4$\cite{Pestun:2007rz,Hama:2012bg},
the deformation breaks the R-symmetry $SU(2)_R$ to $U(1)$.

We can also construct the supersymmetric completion of
the Fayet-Iliopoulos term
${\cal L}=\zeta \tr D_3$
as a special SUSY Chern-Simons action
with the prepotential
\begin{equation}
{\cal F}=ir\zeta(\phi^{(1)})^2\tr\phi.
\end{equation}
Namely, we can regard the Fayet-Iliopoulos parameters
as the coefficients of the linear terms in the prepotential.

Real mass parameters $\mu_n$ for hypermultiplets,
which are associated with global symmetries,
are again introduced by using the constant vector multiplet
(\ref{constantround}).
Let $T_n$ be the generators of the global symmetries associated with
the real mass parameters $\mu_n$.
We weakly gauge $T_n$, and give the expectation values to the corresponding
vector multiplets.
This is realized by shifting the component fields of vector multiplets
in the action 
(\ref{hyperaction})
according to
\begin{equation}
{\cal V}\rightarrow {\cal V}+\mu_n{\cal V}^{(1)}T_n.
\label{realmassshift}
\end{equation}

The vector multiplet ${\cal V}^{(1)}$ is essentially the same as
the central charge vector multiplet introduced in \cite{Kugo:2000hn,Kugo:2000af}.
The real mass parameters
in (\ref{realmassshift})
determine the cenrtal charges
of hypermultiplets.
The Yang-Mills kinetic term is also regarded as the coupling
of the central charge vector multiplet
to the instanton current $j\propto *\tr(F\wedge F)$.
The Yang-Mills coupling constant is a kind of real mass parameters
determining the central charge of instantons.

Let us consider a
theory consisting of a vector multiplet and
an adjoint hypermultiplet.
In the flat background
the global symmetry
$SU(2)_R\times SU(2)_F$ is enhanced to $SO(5)_R$,
and the theory is invariant under ${\cal N}=2$ supersymmetry.
A similar enhancement occurs in ${\bm S}^5$.
In this case, however, a non-trivial mass deformation
is needed
to obtain enhanced supersymmetry\cite{Kim:2012av}.
The mass parameter $\mu_F$ associated with the $SU(2)_F$ flavor symmetry is
introduced by the shift
\begin{equation}
{\cal V}\rightarrow {\cal V}+\mu_F{\cal V}^{(1)}\tau'_3,
\label{su2fmass}
\end{equation}
where $\tau_3'$ is the Cartan generator of $SU(2)_F$.
This mass parameter is related to the deformation parameter $\Delta$
in \cite{Kim:2012av} by $\Delta=1/2+i\mu_F r$.
The supersymmetry enhancement to ${\cal N}=2$ occurs at $\mu_F=\pm\mu_{\rm crit}$
($\mu_{\rm crit}=i/(2r)$).

\subsection{6d interpretation}
As is argued in \cite{Kim:2012av},
supersymmetry on ${\bm S}^5$ can be derived from
that in six-dimensional manifold ${\bm S}^5\times{\bm R}$,
and the enhancement of supersymmetry at the critical points $\mu_F=\pm\mu_{\rm crit}$
is clearly explained from this perspective.
Let us look at this reduction in detail because
this is quite useful when we consider squashing in the next section.

We take the following representation of
6d Dirac matrices.
\begin{equation}
\Gamma^\mu=\left(\begin{array}{cc}
& \gamma^\mu \\
\gamma^\mu
\end{array}\right)\quad(\mu=1,\ldots,5),\quad
\Gamma^6=\left(\begin{array}{cc}
& -i{\bm1}_4 \\
i{\bm1}_4
\end{array}\right),\quad
\Gamma^7=\left(\begin{array}{cc}
{\bm1}_4 \\
& -{\bm1}_4
\end{array}\right).
\end{equation}
We use $M,N,\ldots=1,\ldots,6$ for 6d vector indices,
and assign $12345$ to ${\bm S}^5$ and $6$ to ${\bm R}$.
The ${\cal N}=(1,0)$ superconformal symmetry in 6d is described by
parameters $\epsilon^{(6)}$ and $\kappa^{(6)}$ which have
positive and negative $\Gamma^7$ chirality, respectively.
They satisfy
the six-dimensional Killing equation
\begin{equation}
D_M\epsilon^{(6)}=\Gamma_M\kappa^{(6)}.
\label{kilingeq6d}
\end{equation}
We take the ansatz for the spinors
\begin{equation}
\epsilon^{(6)}=\left(\begin{array}{c}
\epsilon\\
0
\end{array}\right),\quad
\kappa^{(6)}=\left(\begin{array}{c}
0 \\
\kappa
\end{array}\right),
\label{spinor6d}
\end{equation}
where $\epsilon$ and $\kappa$ are the 5d spinors
satisfying the 5d Killing equation (\ref{killing5d}).
We have not yet fixed the normalization of $\epsilon$ and $\kappa$,
which may depend on the coordinate $t\equiv x^6$ along ${\bm R}$.
The relation (\ref{kappas5ande}) gives
\begin{equation}
\kappa^{(6)}=-\frac{1}{2r}\GG\Gamma^6\epsilon^{(6)}.
\label{kappa6d}
\end{equation}
(\ref{kilingeq6d})
is automatically satisfied 
by (\ref{spinor6d})
for $M=1,2,3,4,5$.
Combining (\ref{kilingeq6d}) with $M=6$ and
(\ref{kappa6d}) we obtain
\begin{equation}
\partial_6\epsilon^{(6)}=-\frac{1}{2r}\GG\epsilon^{(6)}.
\end{equation}
This equation determines the $t$ dependence of
$\epsilon$ and $\kappa$.
Because of this non-trivial $t$ dependence
we cannot impose the periodic boundary condition when we
compactify ${\bm R}$ to ${\bm S}^1$.
Instead,
we use the twisted boundary condition
\begin{equation}
\Phi(t+\beta)=\exp\left(-\frac{\beta}{2r}\tau_3\right)\Phi(t),
\label{twistedbc}
\end{equation}
where $\Phi$ is an arbitrary field in the 6d theory,
including $\epsilon$ and $\kappa$.
$\tau_3$ is the Cartan generator of the $SU(2)_R$ symmetry
of the 6d ${\cal N}=(1,0)$ theory.
The Killing spinor satisfies this boundary condition only when
$\epsilon$ satisfies (\ref{s5dircond}).
This is an explanation for (\ref{s5dircond})
in the context of compactification.

The symmetry enhancement in a theory with one adjoint hypermultiplet
at the critical values of
the mass parameter is explained as follows.
Let us start from ${\cal N}=(2,0)$ theory
in 6d, which has $SO(5)_R$ symmetry.
The SUSY
parameters $\epsilon^{(6)}$ and $\kappa^{(6)}$
belong to ${\bm 4}$ of $SO(5)_R$.
The $SU(2)_R$ symmetry of ${\cal N}=(1,0)$ theory is a subgroup of
this $SO(5)_R$, and its centralizer is the flavor group
$SU(2)_F$.
We denote the Cartan generators of $SU(2)_R$ and $SU(2)_F$ by
$\tau_3$ and $\tau_3'$, respectively.
We generalize the twisted boundary condition
(\ref{twistedbc}) by replacing $\tau_3$ by $\tau_3-2ir\mu_F\tau_3'$.
$\mu_F$ is nothing but the mass parameter
in (\ref{su2fmass}).
The condition (\ref{s5dircond}) for preserved SUSY
(\ref{s5dircond}) is replaced by
\begin{equation}
(\tau_3-2ir\mu_F\tau_3')\epsilon=\GG\epsilon.
\end{equation}
For generic $\mu_F$, this condition is satisfied
by a quarter of $\epsilon$, and the preserved SUSY
in 5d is rigid ${\cal N}=1$,
while at the critical values $\mu_F=\pm\mu_{\rm crit}$,
the number of preserved SUSY is doubled.

\section{Squashing}\label{squashing.sec}
\subsection{Supersymmetry}
The squashing of ${\bm S}^5$ can be realized
by a simple modification of
the boundary condition (\ref{twistedbc}).
We consider the boundary condition
\begin{equation}
\Phi(t+\beta)=\exp\left[-\frac{\beta}{2r}((1+\alpha)\tau_3+iuJ)\right]\Phi(t),
\label{twistedbc2def}
\end{equation}
where $J$ is the shift along the Hopf fiber of ${\bm S}^5$
normalized by $e^{2\pi iJ}=1$.
As we will explicitly show shortly,
this gives squashed sphere
(\ref{squashedmetric}) after the dimensional reduction.
The parameter $u$ is related to $v$ in (\ref{squashedmetric})
by
\begin{equation}
v^2=1+u^2.
\end{equation}
The parameter $\alpha$ should be chosen so that there exist preserved SUSY.

The relation (\ref{s5dircond}) is a condition
for $\epsilon$ at each $t$-slice,
and we assume that the change of the boundary condition
does not affect this relation.
Then, the boundary condition
(\ref{twistedbc2def})
implies
\begin{equation}
\alpha\epsilon
=-iu\GG J\epsilon.
\label{n3cond}
\end{equation}

The introduction of the generator $J$ in
the boundary condition breaks $SO(6)_{\rm iso}$ to
$SU(3)\times U(1)$.
Correspondingly, Killing spinors in ${\bm4}+\ol{\bm4}$ split into
${\bm 3}_{+1}+\ol{\bm 3}_{-1}$ and ${\bm 1}_{-3}+{\bm 1}_{+3}$.
If we set $\alpha=-iu$, the condition
(\ref{n3cond}) admits the parameters in the representation
${\bm 3}_{+1}+\ol{\bm 3}_{-1}$.
We call this unbroken supersymmetry ${\cal N}=3/4$.
If we set $\alpha=3iu$, only ${\bm 1}_{-3}+{\bm 1}_{+3}$ are
preserved, and we call this
${\cal N}=1/4$.

\subsection{Killing spinors}\label{killing.sec}
The metric of ${\bm S}^5\times{\bm R}$ is
\begin{equation}
ds^2_{{\bm S}^5\times{\bm R}}=ds_{\bm{CP}^2}^2+(\ol e^5)^2+(\ol e^6)^2.
\end{equation}
The first two terms are the metric of the round ${\bm S}^5$ with radius $r$
in the form of Hopf fibration,
and the last term is the metric of ${\bm R}$.
We introduce coordinates $t$ and $\psi$ along ${\bm R}$ and Hopf fibers,
respectively, and use the local orthonormal frame
\begin{equation}
e^m\quad(m=1,2,3,4),\quad
\ol e^5=d\ol\psi+V,\quad
\ol e^6=dt.
\end{equation}
$e^m$ are the vielbein in the base $\bm{CP}^2$.
In this subsection we use bars to mean the original coordinate system.
We will later introduce a slanted coordinate system in the $56$ plane,
which is convenient for the dimensional reduction.
$V$ is a differential on $\bm{CP}^2$,
which depends on the choice of the coordinate $\psi$.
Its exterior derivative is proportional to the Kahler form $I$ on $\bm{CP}^2$;
\begin{equation}
dV=-\frac{2}{r}I
=-\frac{1}{r}I_{mn}e^m\wedge e^n=\frac{2}{r}(e^1\wedge e^2+e^3\wedge e^4).
\end{equation}
In the small radius limit $\beta\rightarrow0$
almost all modes become infinitely massive.
For modes remaining light
we can replace the twisted boundary condition
(\ref{twistedbc2def}) by the differential equation
\begin{equation}
(\partial_t+u\partial_{\ol\psi})\Phi=-\frac{1+\alpha}{2r}\tau_3\Phi,
\label{slantcond}
\end{equation}
where we used $J=-2ir\partial_{\ol\psi}$.
Note that $\partial_t$ and $\partial_{\ol\psi}$ represents
simple partial derivatives with respect to $t$ and $\ol\psi$,
respectively, and they do not contain the vielbein
unlike $D_\mu$, which has index of the orthonormal frame.

To perform the dimensional reduction
by the condition
(\ref{slantcond}),
it is convenient to introduce
the slanted coordinate
\begin{equation}
\psi=\ol\psi+ut.
\end{equation}
With this coordinate, the compactification is simply represented by
$(t,\psi)\sim (t+\beta,\psi)$.
The metric in the new coordinate system is
\begin{align}
ds_{S^5\times R}^2
&=ds_{\bm{CP}^2}^2
+\frac{1}{v^2}(d\psi+V)^2
+\left(vdt+\frac{u}{v}(d\psi+V)\right)^2.
\end{align}
If we neglect the last term in the metric,
we obtain the squashed ${\bm S}^5$
in (\ref{squashedmetric}).
We introduce the following 6d orthonormal frame
for the slanted coordinate system:
\begin{equation}
e^{m},\quad
e^{5}=\frac{1}{v}(d\psi+V)
,\quad
e^{6}
=vdt+\frac{u}{v}(d\psi+V).
\end{equation}
$e^m$ are the same as before, but $e^5$ and $e^6$
are related to $\ol e^5$ and $\ol e^6$ by
\begin{equation}
\left(\begin{array}{rr}
e^{5} \\
e^{6}
\end{array}\right)
=
\left(\begin{array}{rr}
\frac{1}{v} & -\frac{u}{v} \\
\frac{u}{v} & \frac{1}{v}
\end{array}\right)
\left(\begin{array}{rr}
\ol e^{5} \\
\ol e^{6}
\end{array}\right).
\end{equation}
In this new coordinate system, the constraint
(\ref{slantcond}) becomes
\begin{equation}
D_6=\frac{1}{v}\partial_t=-\frac{1+\alpha}{2rv}\tau_3.
\label{tderslant}
\end{equation}
Note that $\partial_t$ is the simple partial derivative
with respect to $t$, while $D_6$
is the $6$-th component of the covariant derivative
in the orthonormal frame.

To derive the Killing equation on the squashed sphere,
we need to rewrite the 6d Killing equation
(\ref{kilingeq6d}) in terms of 5d language.
The 6d spin connection in the slanted frame has the components
\begin{equation}
\Omega^{mn}=\omega_{\bm{CP}^2}^{mn}+\frac{1}{rv}I_{mn}(e^5+ue^6),\quad
\Omega^{m5}=\frac{1}{rv}I_{mn}e^n,\quad
\Omega^{m6}=\frac{u}{rv}I_{mn}e^n,
\end{equation}
and
the 6d covariant derivative becomes
\begin{align}
D_m^{(6)}
&=\partial_m
+\frac{1}{2}\Omega_m{}^{\mu\nu}S_{\mu\nu}
-\frac{u}{rv}I_{mn}S_{n6},
\label{d6m}\\
D_5^{(6)}
&=v\partial_\psi
-\frac{u}{v}\partial_t
+\frac{1}{2}\Omega_5{}^{\mu\nu}S_{\mu\nu},
\label{d65}\\
D_6^{(6)}
&=\frac{1}{v}\partial_t
+\frac{u}{2rv}I^{\mu\nu}S_{\mu\nu},
\label{d66}
\end{align}
where $S_{MN}$ are spin operators.
For the spinor representation $S_{MN}=(1/2)\Gamma_{MN}$.
Among the components $\Omega_L{}^{MN}$ ($L,M,N=1,\ldots,6$)
of 6d spin connection,
$\Omega_\lambda{}^{\mu\nu}$ ($\lambda,\mu,\nu=1,\ldots,5$)
are identified with the components of
the spin connection on the squashed ${\bm S}^5$,
and we include them in the definition of the
5d covariant derivative.
The $t$ derivatives
are rewritten according to (\ref{tderslant}).
Then the second term in (\ref{d65}) becomes
\begin{equation}
\frac{u(1+\alpha)}{2rv}\tau_3 e^5.
\end{equation}
This can be regarded as a background $SU(2)_R$ gauge field, and
we also include this in the 5d covariant derivative.
As the result, the explicit form of the 5d covariant derivative is
\begin{equation}
D^{(5)}=d-i[A,*]+\frac{1}{2}\Omega_{\mu\nu}[S_{\mu\nu},*]+\frac{u(1+\alpha)}{2rv}e^5[\tau_3,*],
\label{covdeldef}
\end{equation}
where we introduced gauge connection $A$, which has not been
taken into account
in the dimensional reduction.
The 6d covariant derivatives are rewritten as
\begin{align}
D_\mu^{(6)}
&=D_\mu^{(5)}
-\frac{u}{rv}I_{\mu\nu}S_{\nu 6},
\\
D_6^{(6)}
&=-\frac{(1+\alpha)}{2rv}\tau_3
+\frac{u}{2rv}I^{\mu\nu}S_{\mu\nu}.
\end{align}
With these relations and the explicit representation of the 6d Dirac matrices,
we can rewrite the 6d Killing equation (\ref{kilingeq6d}) as
\begin{align}
&D_\mu^{(5)}\epsilon
-\frac{iu}{2rv}I_{\mu\nu}\gamma_{\nu}\epsilon=\gamma_\mu\kappa,
\\
&-\frac{(1+\alpha)}{2rv}\tau_3\epsilon
+\frac{u}{2rv}\sla I\epsilon=-i\kappa,
\label{kappaeps}
\end{align}
and (\ref{kappa6d}) as
\begin{equation}
\kappa
=-\frac{i}{2vr}\tau_3(1+iu\gamma_5)\epsilon.
\end{equation}
By eliminating $\kappa$ from these equations we obtain
the differential equation
\begin{align}
D^{(5)}_\mu\epsilon
=
-\frac{i(1+\alpha)}{2rv}\tau_3\gamma_{\mu}\epsilon
+\frac{iu}{4rv}(
3\gamma_\mu\sla I
-\sla I\gamma_\mu
)\epsilon,
\label{epseq1}
\end{align}
and
the algebraic equation
\begin{equation}
\alpha\epsilon
=
iu\gamma_5\epsilon
+u\tau_3\sla I\epsilon.
\label{epseq2}
\end{equation}
The latter
imposes a condition
on the components of the spinor $\epsilon$ at every point.
This reduces the number of independent components
to six for ${\cal N}=3/4$ and two for ${\cal N}=1/4$.

\subsection{Actions and transformation laws}
Once we have obtained the equations that $\epsilon$ satisfies
it is not difficult to obtain the SUSY
actions and transformation laws by
Noether procedure.
We show only the results.

The transformation laws of vector multiplets are
\begin{align}
\delta \phi&=i(\epsilon\lambda),\nonumber\\
\delta A_\mu&=-(\epsilon\gamma_{\mu}\lambda),\nonumber\\
\delta\lambda
&=
-\sla F\epsilon
+i(\sla D\phi)\epsilon
+iD_a\tau_a\epsilon
+\frac{(1+\alpha)}{rv}\tau_3\epsilon\phi
-\frac{2u}{rv}\sla I\epsilon\phi,
\nonumber\\
\delta D_a
&=i(\epsilon\tau_a\gamma^\mu D_\mu\lambda)
-i(\epsilon\tau_a[\phi,\lambda])
+\frac{(1+\alpha)}{2rv}(\epsilon\tau_3\tau_a\lambda)
-\frac{u}{2rv}(\epsilon\sla I\tau_a\lambda).
\label{deltains}
\end{align}
The SUSY Chern-Simons action is
\begin{align}
{\cal L}_{\rm CS}=&{\cal F}_{\alpha\beta\gamma}\bigg[\left(
\frac{i}{24}\epsilon^{\lambda\mu\nu\rho\sigma}A^\alpha_\lambda F^\beta_{\mu\nu}F^\gamma_{\rho\sigma}
+\cdots\right)
+\frac{i}{4}(\lambda^\alpha\sla F^\beta\lambda^\gamma)
+\frac{1}{4}\phi^\alpha F^\beta_{\mu\nu}F^{\gamma\mu\nu}
+\frac{u}{2rv}\phi^\alpha\phi^\beta I_{\mu\nu}F^{\gamma\mu\nu}
\nonumber\\&
-\frac{1}{4} D_a^\alpha(\lambda^\beta\tau_a\lambda^\gamma)
-\frac{1}{2}\phi^\alpha(\lambda^\beta\sla D\lambda^\gamma)
-\frac{1}{2}\phi^\alpha\lambda^\beta[\lambda,\phi]^\gamma
+\frac{iu}{4rv}\phi^\alpha(\lambda^\beta\sla I\lambda^\gamma)
\nonumber\\&
+\frac{1}{2}\phi^\alpha D_\mu\phi^\beta D^\mu\phi^\gamma
-\frac{1}{2}\phi^\alpha D_a^\beta D_a^\gamma
+\frac{1}{r^2v^2}\left(\frac{2}{3}+\frac{(1+\alpha)^2}{6}+2u^2\right)\phi^\alpha\phi^\beta\phi^\gamma\bigg].
\label{csonsquashed}
\end{align}
The Yang-Mills action and the Fayet-Iliopoulos action
are obtained by taking the prepotential
\begin{equation}
{\cal F}=\frac{1}{2g_{\rm YM}^2}\phi^{(1)}\tr(\phi^2)+\frac{irv}{1+\alpha}\zeta(\phi^{(1)})^2\tr\phi,
\label{ymandfi}
\end{equation}
where
the SUSY preserving constant vector multiplet on the squashed sphere
is
\begin{equation}
{\cal V}^{(1)}
=(\phi^{(1)},A^{(1)},\lambda_I^{(1)},D_a^{(1)})
=\left(1,u e^5,0,\frac{i(1+\alpha)}{rv}\delta_{a3}\right).
\label{squashedphi1}
\end{equation}
The Yang-Mills action
corresponding to the first term in (\ref{ymandfi})
is
\begin{align}
{\cal L}_{\rm YM}
&=\frac{1}{g_{\rm YM}^2}\tr\bigg[\frac{1}{4}F_{\mu\nu}F^{\mu\nu}
+\frac{iu}{8}\epsilon^{5\mu\nu\rho\sigma}F_{\mu\nu}F_{\rho\sigma}
-\frac{1}{2}D_aD_a
-\frac{i(1+\alpha)}{rv}\phi D_3
\nonumber\\&
-\frac{1}{2}\lambda\sla D\lambda
-\frac{iu}{4vr}\lambda\sla I\lambda
-\frac{i(1+\alpha)}{4rv}\lambda\tau_3\lambda
+\frac{1}{2}\lambda[\phi,\lambda]
\nonumber\\&
+\frac{1}{2}D_\mu\phi D^\mu\phi
+\left(\frac{2}{r^2}+\frac{(1+\alpha)^2}{2r^2v^2}\right)\phi^2
\bigg].
\label{ymaction14}
\end{align}
The Fayet-Iliopoulos action
corresponding to the second term in (\ref{ymandfi})
is
\begin{align}
{\cal L}_{\rm FI}
&=
\zeta\tr\bigg[D
-\frac{iu}{2(1+\alpha)}I_{\mu\nu}F^{\mu\nu}
+\frac{4u^2}{rv(1+\alpha)}A_5
\nonumber\\&
+\frac{i}{rv(1+\alpha)}\left(
2
+(1+\alpha)^2
+2u^2\right)\phi\bigg].
\end{align}

The transformation laws for hypermultiplets are
\begin{align}
\delta q_i=&i(\epsilon\rho_i\psi),\nonumber\\
\delta\psi=&
i\ol\rho_i(\sla D q_i) \epsilon
+\frac{3(1+\alpha)}{2rv}\ol\rho_i\tau_3\epsilon q_i
-\frac{2u}{rv}\ol\rho_i\sla I\epsilon q_i
-i\ol\rho_{i}\epsilon [\phi,q_i].
\end{align}
The kinetic action of hypermultiplets is
\begin{align}
{\cal L}_{\rm hyper}
&=\frac{1}{2}\psi\sla D\psi
+\frac{iu}{4vr}\psi\sla I\psi
+\frac{1}{2}\psi[\phi,\psi]
+\psi\ol\rho_i[\lambda,q_i]
\nonumber\\&
+\frac{1}{2}D_\mu q_i D^\mu q_i
+\left(\frac{2}{r^2}
-\frac{(1+\alpha)^2}{8r^2v^2}\right)q_iq_i
\nonumber\\&
+\frac{1}{2}(\tau_a)_{ij}q_i[D_a,q_j]
-\frac{1}{2}[\phi,q_i][\phi,q_i].
\label{hyperactionx}
\end{align}

By shifting the vector multiplet fields in (\ref{hyperactionx})
by ${\cal V}\rightarrow{\cal V}+\mu_{\rm re}{\cal V}^{(1)}$,
the following real mass terms arise.
\begin{equation}
{\cal L}_{\rm hyper}^{\rm real\ mass}=\frac{1}{2}\psi[\mu_{\rm re},\psi]
-\frac{iu}{2}\psi\gamma_5[\mu_{\rm re},\psi]
+\frac{1+u^2}{2}[q_i,\mu_{\rm re}][\mu_{\rm re},q_i]
+\frac{i(1+\alpha)}{2rv}(\tau_3)_{ij}q_i[\mu_{\rm re},q_j].
\label{realmass}
\end{equation}

\subsection{More actions in ${\cal N}=1/4$}
When $\alpha=3iu$, the relation (\ref{epseq2}) implies
$\tau_3\sla I\epsilon=2i\epsilon$,
and
there is another
SUSY preserving constant vector multiplet
in ${\cal N}=1/4$ theory.
\begin{equation}
{\cal V}^{(2)}=(\phi^{(2)},A^{(2)},\lambda_I^{(2)},D_a^{(2)})
=\left(0,e^5,0,-\frac{4}{rv}\delta_{a3}\right).
\label{newconst}
\end{equation}
With this multiplet we can construct the following quadratic action
of vector multiplets
corresponding to the prepotential ${\cal F}=(1/2)\phi^{(2)}\tr\phi^2$.
\begin{align}
{\cal L}_1
&=\tr\left[
\frac{i}{8}\epsilon^{5\mu\nu\rho\sigma}
F_{\mu\nu}F_{\rho\sigma}
-\frac{i}{2rv}\lambda\sla I\lambda
-\frac{1}{rv}\phi I^{\mu\nu}F_{\mu\nu}
-\frac{4u}{r^2v^2}\phi\phi
+\frac{1}{rv}\lambda\tau_3\lambda
+\frac{4}{rv}\phi D_3\right].
\end{align}
This is not independent observable from the Yang-Mills action
(\ref{ymaction14})
in the sense that ${\cal L}_1$ and ${\cal L}_{\rm YM}$
are proportional to each other
as $Q$-cohomology classes.
Let us define $Q_\pm$ by
\begin{equation}
\delta(\eta\varepsilon_\pm)=\eta Q_\pm,
\label{qpmdef}
\end{equation}
where the left hand side stands for thetransformation with the
Grassmann-odd parameter $\eta\varepsilon_\pm$.
$\eta$ is a constant Grassmann-odd number and
$\varepsilon_\pm$ are the bosonic Killing spinors satisfying
\begin{equation}
\tau_3\varepsilon_\pm=\pm\varepsilon,\quad
\varepsilon_+^\dagger\varepsilon_+
=\varepsilon_-^\dagger\varepsilon_-=1.
\end{equation}
We can show
\begin{align}
{\cal L}_{\rm YM}+\frac{i(1+iu)}{g_{\rm YM}^2}{\cal L}_1
=& \frac{1}{g_{\rm YM}^2}Q_+\tr\left[\frac{1}{2}(Q_+\lambda)^\dagger\lambda+\frac{2(1-iu)}{rv}
\varepsilon_+^\dagger\lambda\phi\right].
\label{ymp1}
\end{align}
Note that the bosonic part of (\ref{ymp1})
is not positive definite, and we cannot use this
to localize the path integral.

We can introduce mass terms of hypermultiplets using (\ref{newconst}).
This can be regarded as the imaginary counterpart of
the real mass terms (\ref{realmass}).
By shifting the vector multiplets in (\ref{hyperactionx}) by
\begin{equation}
{\cal V}\rightarrow{\cal V}+\mu_{\rm re}{\cal V}^{(1)}+(\mu_{\rm im}-u\mu_{\rm re}){\cal V}^{(2)},
\end{equation}
we obtain the mass terms
\begin{align}
{\cal L}_{\rm hyper}^{\rm complex\ mass}
=&\frac{1}{2}[q_i,\mu_{\rm re}][\mu_{\rm re},q_i]
+\frac{1}{2}[q_i,\mu_{\rm im}][\mu_{\rm im},q_i]
+\frac{1}{2}\psi[\mu_{\rm re},\psi]
-\frac{i}{2}\psi\gamma_5[\mu_{\rm im},\psi]
\nonumber\\&
+\frac{i(1+iu)}{2rv}(\tau_3)_{ij}q_i[\mu_{\rm re},q_i]
-\frac{2}{rv}(\tau_3)_{ij}q_i[\mu_{\rm im},q_i].
\end{align}

\subsection{${\cal N}=1/2$ and ${\cal N}=3/2$}
In a gauge theory with a single adjoint hypermultiplet,
the enhancement of supersymmetry occurs just as on the round ${\bm S}^5$.
We turn on the critical value of the $SU(2)_F$ mass parameter by the
shift
\begin{equation}
{\cal V}\rightarrow {\cal V}+\frac{i(1+\alpha)}{2rv}{\cal V}^{(1)}\tau_3'.
\label{shiftto2}
\end{equation}
This corresponds to the modification of the boundary condition
(\ref{twistedbc2def}) to
\begin{equation}
\Phi(t+\beta)=\exp\left[-\frac{\beta}{2r}((1+\alpha)(\tau_3+\tau_3')+iuJ)\right]\Phi(t).
\end{equation}
The Killing equation for the enhanced supersymmetry
is obtained from the 6d ${\cal N}=(2,0)$ supersymmetry.
$\epsilon$ and $\kappa$ are $SO(5)_R$ quartet, and satisfy
equations
(\ref{epseq1}), and (\ref{epseq2})
with $\tau_3$ replaced by $\tau_3+\tau_3'$.
We can see that there are $12$ supercharges for $\alpha=-iu$
and $4$ for $\alpha=3iu$.
We call these supersymmetries ${\cal N}=3/2$ and ${\cal N}=1/2$,
respectively.
To write down the actions and the transformation laws
in $SO(5)_R$ covariant form,
we embed fields and the SUSY parameters of
${\cal N}=1/4$ or $3/4$ theory into $SO(5)_R$ multiplets as
\begin{equation}
q_\alpha=(q_i,q_5)=(q_i,\phi),\quad
\chi_a=\left(\begin{array}{c}
\lambda_I \\
\psi_A
\end{array}\right),\quad
\epsilon_a=\left(\begin{array}{c}
\epsilon_I \\
0
\end{array}\right).
\end{equation}
After the elimination of the auxiliary fields $D_a$,
we obtain the
transformation laws
\begin{align}
\delta q_\alpha=&i(\epsilon\wh\rho_\alpha\chi),\nonumber\\
\delta A_\mu=&-(\epsilon\gamma_{\mu}\chi),\nonumber\\
\delta\psi=&-\sla F\epsilon
+i\wh\rho_\alpha(\sla D q_\alpha) \epsilon
+\frac{1+\alpha}{2rv}\wh\rho_\alpha\epsilon (\tau_3+\tau_3')q_\alpha
\nonumber\\&
+\frac{2}{rv}\wh\rho_\alpha(\tau_3+\tau_3')(1+\alpha)\epsilon q_\alpha
-\frac{2u}{rv}\wh\rho_\alpha\sla I\epsilon q_\alpha
+\frac{i}{2}\wh\rho_{\alpha\beta}\epsilon [q_\alpha,q_\beta],
\end{align}
and the action
\begin{align}
{\cal L}=&\tr\bigg[\frac{1}{4}F_{\mu\nu}F^{\mu\nu}+\frac{iu}{8}\epsilon^{5\mu\nu\rho\sigma}F_{\mu\nu}F_{\rho\sigma}\nonumber\\
&-\frac{1}{2}\chi\sla D\chi
-\frac{iu}{4vr}\chi\sla I\chi
-\frac{i(1+\alpha)}{4rv}\chi(\tau_3+\tau_3')\chi
-\frac{1}{2}\chi\wh\rho_\alpha[\chi,q_\alpha]
\nonumber\\
&+\frac{1}{2}D_\mu q_\alpha D^\mu q_\alpha
-\frac{(1+\alpha)^2}{2r^2v^2}(q_1^2+q_2^2)
+\frac{2}{r^2}q_\alpha q_\alpha
\nonumber\\&
-\frac{1}{4}[q_\alpha,q_\beta][q_\alpha,q_\beta]
-\frac{1+\alpha}{3rv}\epsilon_{12\alpha\beta\gamma}[q_\alpha,q_\beta]q_\gamma
\bigg],
\end{align}
where $SO(5)_R$ Dirac matrices $\wh\rho_\alpha$ are defined in the appendix.

\section{Discussion}
In this paper, we constructed SUSY transformation laws
and SUSY actions in the $SU(3)\times U(1)$ symmetric squashed
five-sphere.
An important task we should try next is to compute
the partition function.
Although the instanton contribution has not yet been computed
even for the round sphere,
it should be possible to compute the perturbative sector
of the partition function
for the squashed ${\bm S}^5$.

In the case of the round sphere,
the saddle points in the perturbative sector are parameterized by
the constant expectation values of the scalar fields $a^\alpha$ in
the vector multiplets,
and the on-shell action is obtained by substituting
the constant vector multiplets
${\cal V}^\alpha=a^\alpha{\cal V}^{(1)}$
to the action.
This is also the case for the squashed ${\bm S}^5$.
The classical Lagrangian density at the saddle point is
\begin{equation}
{\cal L}=\frac{1}{v^2r^2}\left[4+4(1+\alpha)^2+12u^2-8iu^3\right]
{\cal F}(a^\alpha).
\end{equation}
If we multiply the volume of the squashed sphere
$\pi^3r^5/v$
and set $\alpha=-iu$ and $\alpha=3iu$, we obtain the classical action
for ${\cal N}=3/4$ and ${\cal N}=1/4$
\begin{equation}
S_{{\cal N}=\frac{3}{4}}=(2\pi r)^3\left(\frac{1+iu}{v}\right)^{-1}
{\cal F}(a^\alpha),\quad
S_{{\cal N}=\frac{1}{4}}=(2\pi r)^3\left(\frac{1+iu}{v}\right)^3
{\cal F}(a^\alpha).
\label{clasactuon}
\end{equation}
Interestingly, in both cases the classical action depends
on the squashing parameter through $(1+iu)/v$.
This is similar to the case of the $SU(2)\times U(1)$ invariant squashing
of ${\bm S}^3$\cite{Imamura:2011wg}.
Of course we cannot conclude whether the partition function
depends on the squashing parameter
until we compute the one-loop contribution
because the dependence of the
classical action may be absorbed in the normalization
of the integration variables $a^\alpha$.

When we compute the partition function by localization,
we need to choose one supercharge $Q$.
Let us consider the ${\cal N}=1/4$ case.
In this case $Q$ is a linear combination of
$Q_\pm$ defined by (\ref{qpmdef}).
If we choose $Q=aQ_++bQ_-$, its square is
\begin{equation}
Q^2=2abv\left[-{\cal L}_\psi-\frac{3i}{2r}\tau_3\right]+\mbox{gauge tr.},
\label{n14qsq}
\end{equation}
where ${\cal L}_\psi$ is the Lie derivative along the vector field $\partial_\psi$.
The squashing parameter dependence is factorized
up to the field dependent gauge transformation term,
and we can absorb it by the coefficients $a$ and $b$.
This factorization strongly suggests that the partition function
is independent of the squashing parameter.
Indeed,
in \cite{Kallen:2012cs,Kallen:2012va}
the partition function
is computed
based on the algebra (\ref{n14qsq}),
which is compatible with the contact structure of the
manifold.
It would be possible to apply the method in
\cite{Kallen:2012cs,Kallen:2012va} to ${\cal N}=1/4$ theories
on the squashed sphere.

Another way to
obtain the partition function of ${\cal N}=1/4$ theory
is the direct calculation based on the harmonic expansion
used in \cite{Kim:2012av}.
Because the $Q$-exact terms used in \cite{Kim:2012av}
breaks $SO(6)$ isometry
of the round sphere to $SU(3)\times U(1)$,
the computation in \cite{Kim:2012av}
does not rely on the full $SO(6)$ isometry,
and the extension to the squashed sphere,
which also has $SU(3)\times U(1)$ symmetry, is straightforward.
For vector multiplets, we obtain
\begin{equation}
Z_{\rm vector}^{\rm 1-loop}=\prod_{\alpha\in{\rm root}}
\prod_{k=1}^\infty
\left(k+i\frac{1+iu}{v}\alpha(a)\right)^{k^2+2}
\end{equation}
up to a constant factor.
Therefore, the squashing parameter dependence is
absorbed by the rescaling
\begin{equation}
\frac{1+iu}{v}a^\alpha\rightarrow a^\alpha
\label{rescaling}
\end{equation}
of the
integration variables $a^\alpha$.
This is nothing but the rescaling needed to absorb
the squashing parameter dependence of the
classical action $S_{{\cal N}=\frac{1}{4}}$ in (\ref{clasactuon}).
After the rescaling
(\ref{rescaling}),
the expression of the partition function
becomes identical to that of the round ${\bm S}^5$.
Although we have not computed the partition function
of hypermultiplets,
it seems unlikely to depend on the squashing parameter.

On the other hand,
in the case of ${\cal N}=3/4$,
there is no supercharges compatible with the contact structure.
Namely, there is no supercharge $Q$ such that $Q^2$ generates
shift along the Hopf fiber.
In this case
the partition function may
depend on the squashing parameter.

The situation above is very similar to the 3d case.
On the $SU(2)\times U(1)$ symmetric squashed ${\bm S}^3$,
there are two kinds of supersymmetry.
One is $SU(2)$ singlet supersymmetry.
In this case, the partition function does not depend on the squashing parameter\cite{Hama:2011ea}.
See also \cite{Kallen:2011ny,Ohta:2012ev} for the analysis based on the
contact structure.
The other is $SU(2)$ doublet supersymmetry.
In this case, the partition function depends on the
squashing parameter\cite{Imamura:2011wg}
just as in the case of the ellipsoidal deformation\cite{Hama:2011ea}.

It is an interesting problem whether the partition function of
a 5d ${\cal N}=3/4$ theory depends on the squashing parameter.
We hope to return to this problem in the near future.

\section*{Acknowledgments}
The author would like to thank Taichiro Kugo for useful information.
Y.~I. is partially supported by Grant-in-Aid for Scientific Research
(C) (No.24540260), Ministry of Education, Science and Culture, Japan.

\appendix
\section{Appendix}
\subsection{Conventions for $SU(2)$ and $SO(5)$}
$SU(2)_R$ generators $(\tau_a)_I{}^J$ and $SU(2)_F$ generators $(\tau_a')_A{}^B$ ($a=1,2,3$) are defined by
\begin{equation}
\tau_1=\tau_1'=\sigma_x,\quad
\tau_2=\tau_2'=\sigma_y,\quad
\tau_3=\tau_3'=\sigma_z,
\end{equation}
where $\sigma_i$ are Pauli matrices:
\begin{equation}
\sigma_x=\left(\begin{array}{cc}
0 & 1 \\
1 & 0
\end{array}\right),\quad
\sigma_y=\left(\begin{array}{cc}
0 &-i \\
i & 0
\end{array}\right),\quad
\sigma_z=\left(\begin{array}{cc}
1 & 0 \\
0 & -1
\end{array}\right).
\end{equation}
We follow the NW-SE rule for the contraction of $SU(2)$ indices,
and when we need to raise an index, we use $\epsilon$ tensor
with components $\epsilon^{12}=-\epsilon^{21}=1$.
The $SU(2)$ invariant product of two doublets are defined by
\begin{equation}
XY=X^IY_I=(\epsilon^{IJ}X_J)Y_I.
\end{equation}
$SU(2)_R\times SU(2)_F$ invariant tensor
$(\rho_i)_I{}^A$
and $(\ol\rho_i)_A{}^I$ ($i=1,2,3,4$) are defined by
\begin{equation}
\rho_i=(\sigma_x,\sigma_y,\sigma_z,-i{\bm1}_2),\quad
\ol\rho_i=(\sigma_x,\sigma_y,\sigma_z,i{\bm1}_2).
\end{equation}

$SO(5)_R$ Dirac matrices $\wh\rho_\alpha$ ($\alpha=1,2,3,4,5$)
are defined by
\begin{equation}
\wh\rho_i=\left(\begin{array}{cc}
& \rho_i \\
\ol\rho_i
\end{array}\right),\quad
\wh\rho_5=\left(\begin{array}{cc}
{\bm1}_4 \\
& -{\bm1}_4
\end{array}\right).
\end{equation}
Let $\chi_a$ ($a=1,2,3,4$) be the $SO(5)_R$ quartet
consists of $SU(2)_R$ doublet $\lambda_I$
and $SU(2)_F$ doublet $\psi_A$,
\begin{equation}
\chi_a=(\chi_1,\chi_2,\chi_3,\chi_4)
=(\lambda_1,\lambda_2,\psi_1,\psi_2),
\end{equation}
and $\chi'_a$ be defined from $\lambda'_I$ and $\psi'_A$ in the same way.
The $SO(5)_R$-invariant product of these two $SO(5)_R$ quartets
is defined by
\begin{equation}
\chi\chi'
=\lambda\lambda'-\psi\psi'
=\epsilon^{IJ}\lambda_J\lambda'_I-\epsilon^{IJ}\psi_J\psi'_I
\label{so5rinvprod}
\end{equation}

The spacetime Dirac matrices $\gamma_\mu$
have the same components as $\wh\rho_\alpha$
\begin{equation}
\gamma_1=\wh\rho_1,\quad
\gamma_2=\wh\rho_2,\quad
\gamma_3=\wh\rho_3,\quad
\gamma_4=\wh\rho_4,\quad
\gamma_5=\wh\rho_5.
\end{equation}

$\rho$, $\ol\rho$, $\wh\rho$, and $\gamma$ with multiple indices
represent anti-symmetric products.
For example,
\begin{equation}
\gamma_{\mu\nu}=\frac{1}{2}
(\gamma_\mu\gamma_\nu-\gamma_\nu\gamma_\mu).
\end{equation}
Backslashes represent the contraction with Dirac matrices.
For example
\begin{equation}
\sla F=\frac{1}{2}\gamma^{\mu\nu}F_{\mu\nu}.
\end{equation}
The scalar product of two spinors are defined in the same way as
$SO(5)_R$-invariant product (\ref{so5rinvprod}).

$SO(5)_R$ invariant antisymmetric tensor is defined by
\begin{equation}
\wh\rho^{\alpha\beta\gamma\delta\epsilon}
=\epsilon^{\alpha\beta\gamma\delta\epsilon}{\bm 1}_4.
\end{equation}
With our representation of Dirac matrices,
this has the component $\epsilon^{12345}=-1$.
The spacetime antisymmetric tensor has the same components
with this.



\begin{thebibliography}{99}
\bibitem{Seiberg:1996bd} 
  N.~Seiberg,
  ``Five-dimensional SUSY field theories, nontrivial fixed points and string dynamics,''
  Phys.\ Lett.\ B {\bf 388}, 753 (1996)
  [hep-th/9608111].
\bibitem{Douglas:2010iu} 
  M.~R.~Douglas,
  ``On D=5 super Yang-Mills theory and (2,0) theory,''  JHEP {\bf 1102}, 011 (2011)  [arXiv:1012.2880 [hep-th]].  
\bibitem{Lambert:2010iw} 
  N.~Lambert, C.~Papageorgakis and M.~Schmidt-Sommerfeld,
  ``M5-Branes, D4-Branes and Quantum 5D super-Yang-Mills,''  JHEP {\bf 1101}, 083 (2011)  [arXiv:1012.2882 [hep-th]].  
\bibitem{Hosomichi:2012ek} 
  K.~Hosomichi, R.~-K.~Seong and S.~Terashima,
  ``Supersymmetric Gauge Theories on the Five-Sphere,''
  arXiv:1203.0371 [hep-th].
\bibitem{Kim:2012av} 
  H.~-C.~Kim and S.~Kim,
  ``M5-branes from gauge theories on the 5-sphere,''
  arXiv:1206.6339 [hep-th].
\bibitem{Kim:2012gu} 
  H.~-C.~Kim, S.~-S.~Kim and K.~Lee,
  ``5-dim Superconformal Index with Enhanced En Global Symmetry,''
  arXiv:1206.6781 [hep-th].
\bibitem{Terashima:2012ra} 
  S.~Terashima,
  ``On Supersymmetric Gauge Theories on S$^4$ x S$^1$,''
  arXiv:1207.2163 [hep-th].
\bibitem{Kawano:2012up} 
  T.~Kawano and N.~Matsumiya,
  ``5D SYM on 3D Sphere and 2D YM,''  arXiv:1206.5966 [hep-th].  
\bibitem{Kallen:2012cs} 
  J.~Kallen and M.~Zabzine,
  ``Twisted supersymmetric 5D Yang-Mills theory and contact geometry,''
  JHEP {\bf 1205}, 125 (2012)
  [arXiv:1202.1956 [hep-th]].
\bibitem{Kallen:2012va} 
  J.~Kallen, J.~Qiu and M.~Zabzine,
  ``The perturbative partition function of supersymmetric 5D Yang-Mills theory with matter on the five-sphere,''
  arXiv:1206.6008 [hep-th].
\bibitem{Kallen:2012zn} 
  J.~Kallen, J.~A.~Minahan, A.~Nedelin and M.~Zabzine,
  ``$N^3$-behavior from 5D Yang-Mills theory,''  arXiv:1207.3763 [hep-th].  
\bibitem{Jafferis:2012iv} 
  D.~L.~Jafferis and S.~S.~Pufu,
  ``Exact results for five-dimensional superconformal field theories with gravity duals,''
  arXiv:1207.4359 [hep-th].
\bibitem{Pestun:2007rz} 
  V.~Pestun,
  ``Localization of gauge theory on a four-sphere and supersymmetric Wilson loops,''  Commun.\ Math.\ Phys.\  {\bf 313}, 71 (2012)  [arXiv:0712.2824 [hep-th]].  
\bibitem{Hama:2012bg} 
  N.~Hama and K.~Hosomichi,
  ``Seiberg-Witten Theories on Ellipsoids,''  arXiv:1206.6359 [hep-th].  
\bibitem{Kugo:2000hn} 
  T.~Kugo and K.~Ohashi,
  Prog.\ Theor.\ Phys.\  {\bf 104}, 835 (2000)
  [hep-ph/0006231].
\bibitem{Kugo:2000af} 
  T.~Kugo and K.~Ohashi,
  Prog.\ Theor.\ Phys.\  {\bf 105}, 323 (2001)
  [hep-ph/0010288].
\bibitem{Imamura:2011wg} 
  Y.~Imamura and D.~Yokoyama,
  ``N=2 supersymmetric theories on squashed three-sphere,''
  Phys.\ Rev.\ D {\bf 85}, 025015 (2012)
  [arXiv:1109.4734 [hep-th]].
\bibitem{Hama:2011ea} 
  N.~Hama, K.~Hosomichi and S.~Lee,
  ``SUSY Gauge Theories on Squashed Three-Spheres,''  JHEP {\bf 1105}, 014 (2011)  [arXiv:1102.4716 [hep-th]].  
\bibitem{Kallen:2011ny} 
  J.~Kallen,
  ``Cohomological localization of Chern-Simons theory,''
  JHEP {\bf 1108}, 008 (2011)
  [arXiv:1104.5353 [hep-th]].

\bibitem{Ohta:2012ev} 
  K.~Ohta and Y.~Yoshida,
  ``Non-Abelian Localization for Supersymmetric Yang-Mills-Chern-Simons Theories on Seifert Manifold,''
  arXiv:1205.0046 [hep-th].
\end{thebibliography}
\end{document}